%% file: Zoller_Radcor_LoopFest_2021.tex
\documentclass[submission, Proceedings]{SciPost}

\hypersetup{
    colorlinks,
    linkcolor={red!50!black},
    citecolor={blue!50!black},
    urlcolor={blue!80!black}
}

\usepackage[bitstream-charter]{mathdesign}
\usepackage{enumitem}
\input{macros_OL_2l}

\graphicspath{{./figures/}}

\DeclareSymbolFont{usualmathcal}{OMS}{cmsy}{m}{n}
\DeclareSymbolFontAlphabet{\mathcal}{usualmathcal}

\begin{document}

\begin{center}{\Large \textbf{
Two-loop amplitude generation in OpenLoops
}}\end{center}

\begin{center}
{Stefano Pozzorini}\textsuperscript{1},
{Natalie Sch\"ar}\textsuperscript{2} and
{Max F. Zoller}\textsuperscript{2$\star$}
\end{center}

\begin{center}
{\bf 1} Universit\"at Z\"urich, CH-8057 Z\"urich, Switzerland
\\
{\bf 2} Paul Scherrer Institut, CH-5232 Villigen PSI, Switzerland
\\
* max.zoller@psi.ch
\end{center}

\begin{center}
\today
\end{center}


\definecolor{palegray}{gray}{0.95}
\begin{center}
\colorbox{palegray}{
  \begin{tabular}{rr}
  \begin{minipage}{0.1\textwidth}
    \includegraphics[width=35mm]{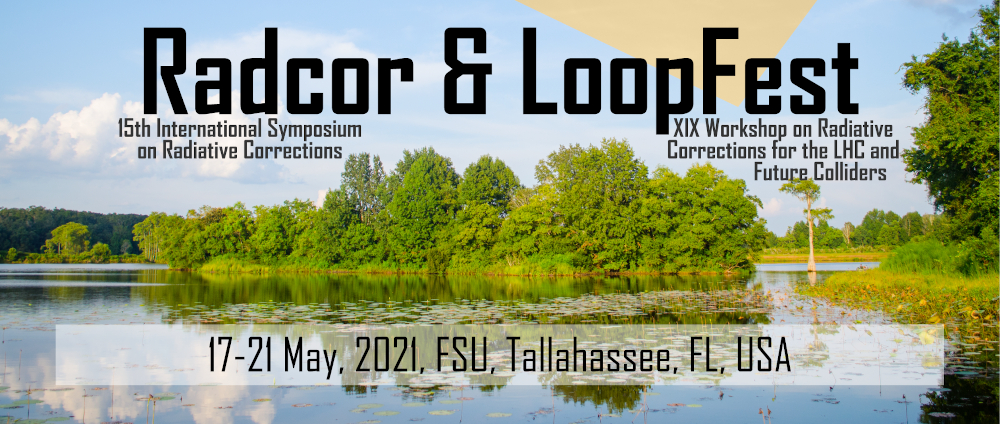}
  \end{minipage}
  &
  \begin{minipage}{0.85\textwidth}
    \begin{center}
    {\it 15th International Symposium on Radiative Corrections: \\Applications of Quantum Field Theory to Phenomenology,}\\
    {\it FSU, Tallahasse, FL, USA, 17-21 May 2021} \\
    \href{https://scipost.org/SciPostPhysProc.202106001}{SciPostPhysProc.202106001}\\
    \end{center}
  \end{minipage}
\end{tabular}
}
\end{center}

\section*{Abstract}
{\bf
Numerical tools, such as OpenLoops, provide NLO scattering amplitudes for a very wide range of hard scattering amplitudes in a fully automated way. In order to match the numerical precision of current and future experiments, however, the higher precision of NNLO calculations is essential, and their automation in a similar tool highly desirable.

In our approach, D-dimensional amplitudes are decomposed into loop-momentum tensor integrals with coefficients constructed in four dimensions and rational terms. We present a fully generic algorithm for the efficient numerical construction of the tensor coefficients, which constitutes an important building block for an automated NNLO tool.
}

\vspace{10pt}
\noindent\rule{\textwidth}{1pt}
\tableofcontents\thispagestyle{fancy}
\noindent\rule{\textwidth}{1pt}
\vspace{10pt}

\section{Introduction}
\label{sec:intro}
Precise Monte Carlo simulations of scattering processes have played a major role in the success of the LHC. The hard scattering amplitudes at the core of these simulations can be obtained by fully automated numerical tools, such as 
\OpenLoops{} \cite{Cascioli:2011va,Buccioni:2017yxi,Buccioni:2019sur}, at tree and one-loop level.
This is sufficient in order to obtain LO and NLO predictions for most processes, but in order to fully exploit the potential of the LHC and future colliders, NNLO predictions are required for a wide range of processes. While dedicated NNLO calculations, which involve two-loop amplitudes, exist for many $2 \to 2$ and a few $2 \to 3$ processes, a fully automated NNLO tool for processes with four, five and possibly more scattering particles would greatly expand the scope of precision phenomenology. 

In the following, we will summarize the building blocks of the NLO \OpenLoops{} program, and describe the ones required for a future NNLO tool. We will then present a major building block for a NNLO \OpenLoops{} program, namely a new algorithm for the numerical construction of two-loop amplitudes in terms of loop-momentum tensor integrals.

\section{Tree-level and one-loop amplitude construction in OpenLoops}
\label{sec:oneloop}
$L$-loop scattering amplitudes are computed as sums of Feynman diagrams $\Gamma$, 
\be
\calM_{L}(\heli) \,=\, \sum\limits_\Gamma \fullamp{L}{\Gamma}(\heli){},
\ee
the amplitudes of which depend on the helicity configuration $\heli$ of the external particles and are factorised into a colour factor and colour-stripped amplitude,
\be
\fullamp{L}{\Gamma}(\heli)
\,=\,
\colfac{L}{\Gamma}\,\amp{L}{\Gamma}(\heli){}. \label{eq:amp_colfact}
\ee
While the colour factors $\colfac{L}{\Gamma}$ are handled algebraically, the colour-stripped amplitudes $\amp{L}{\Gamma}$ are constructed numerically in \OpenLoops{}.

Tree-level diagrams are decomposed into subtrees $w_a$,
represented as blue bubbles in our graphs, which are then constructed through recursion steps,
\be
w^{\alpha}_a \,=\,
\parbox{0.14\textwidth}{\includegraphics[height=10mm]{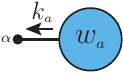}} \,=\,
\parbox{0.14\textwidth}{\includegraphics[height=20mm]{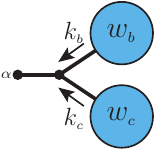}} \,=\,
\f{X_{\beta\gamma}^{\alpha}(k_b,k_c)}{\momk{a}^2-\mass{a}^2}
w^{\beta}_b 
w^{\gamma}_c ,
\ee
from two subtrees with less propagators and a universal function $X$ derived from the Feynman rule of the connecting vertex and adjacent propagator. The denominator contains the mass $\mass{a}$ and momentum $\momk{a}$ of this propagator. The recursion starts
from the external wave functions, and ends in connecting two subtrees which form the full diagram. This recursion is implemented in four dimensions, 
achieving a high level of efficiency through the recycling of
already constructed subtrees in multiple tree-level and loop diagrams.\footnote{Subtrees can be factorised from divergent loop diagrams as well as from the corresponding counterterm diagrams, which allows for their construction in four dimensions, since the sum of these diagrams is finite.} 

Starting from one-loop level, divergences can appear and need to be treated through renormalisation and IR subtraction procedures. In addition, the numerators of Feynman integrals
are constructed in integer dimensions in a numerical tool.
Hence, one-loop amplitudes $\barM_{1}$ in $D$ dimensions are split into an amplitude $\calM_{1}$ constructed from Feynman integrals with four-dimensional numerators and 
a remainder stemming from $(D-4)$-dimensional numerators.
The latter can be fully reconstructed through rational counterterm \cite{Ossola:2008xq,
Draggiotis:2009yb,
Garzelli:2009is,
Pittau:2011qp} 
insertions into tree level amplitudes, which are computed together with the one-loop UV counterterms in the chosen renormalisation scheme as $\calM_{0,\rm{1l-CT}}$.

For a large class of processes the helicity and colour-summed squared tree-level amplitude
\be
\calW_{\ssst{LO}}\,=\,\f{1}{N_{\rm{hcs}}}
\sum\limits_{\heli,\col}
|\calM_{0}(\heli)|^2{},
\label{M2Wtree}
\ee
where $1/N_{\rm{hcs}}$ encodes the average over initial-state helicity and colour d.o.f as well as symmetry factors for identical final-state particles (see \cite{Buccioni:2019sur}),
constitutes the LO contribution of the scattering probability density, while the NLO contribution
is computed from the Born-loop interference
\be
\calW_{\ssst{NLO}}^{\ssst{virtual}}\,=\,\f{1}{N_{\rm{hcs}}}
\sum\limits_{\heli,\col} 
2\,\re \Big[\calM_{0}^*(\heli)\calM_{1}(\heli) + \calM_{0}^*(\heli)\calM_{0,\rm{1l-CT}}(\heli)\Big].
\label{M2Wone}
\ee

The colour-stripped amplitude of a one-loop diagram $\Gamma$ is given by
\be\mathcal{A}_{1,\Gamma} =\,  \vcenter{\hbox{\scalebox{1.}{\includegraphics[height=25mm]{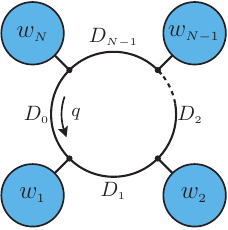}}}} 
= \int\!\rd\momq\; \f{{\rm Tr}\left[\seg_1(q)\!\cdots\!\seg_N({q})\right]}{ 
D_{0}\!\cdots\! D_{N-1}}
\ee
with the integration measure in loop momentum space 
$\int\!\rd\momq = \mu^{2\eps} \int \f{\rd^{^D}\! \bar
q}{(2\pi)^{^D}}$ and 
scalar propagator denominators $D_{a}(q)=(q+p_a)^2-m_a^2$
with mass $m_a$ and external momentum $p_a$.
The numerator factorises into loop segments with at most linear $q$-dependence,
\bea
\seg_a(q)&=&\parbox{0.13\textwidth}{
\includegraphics[height=18mm]{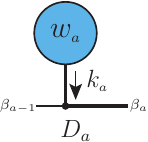}}
=
\{Y_{\sigma}^{a}+ Z_{\nu;\sigma}^{a}\,{q^\nu }
\}\, w^{\sigma}_a ,
\eea
which consist of a loop vertex and propagator encoded in the universal building blocks $Y, Z$ and one or two external sub-trees $w_a$ with external momentum $k_a$. These segments should be understood as matrices with Lorentz or spinor indices $\beta_{a-1}, \beta_a$.

In \OpenLoops{}, the one-loop diagram is cut open at a chosen off-shell propagator $D_0$ and the resulting chain of segments constructed recursively through steps ($k=1,\ldots,N$)
\be
\calN_k(q)
\,=\,\calN_{k-1}(q)\seg_{k}(q)=\,\vcenter{\hbox{\includegraphics[width=0.53\textwidth]{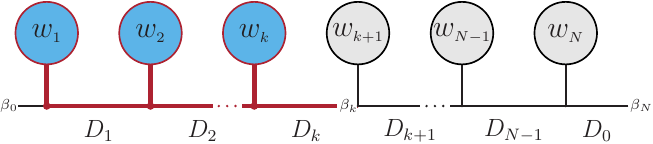}}}
\ee
starting from $\calN_0=1$. The numerator can be
written as
\be
\calN_k(q)=\prod_{i=1}^{k}\seg_i(q) = \sum\limits_{r=0}^k
\calN_{k,\mu_1\ldots\mu_{r}}{q^{\mu_1}\ldots q^{\mu_{r}} },
\ee
and the numerical recursion is implemented at
the level of the tensor coefficients $\calN_{k,\mu_1\ldots\mu_{r}}$, retaining the analytical structure in $q$ throughout the amplitude construction. The tensor integrals in the resulting amplitude
\be\mathcal{A}_{1,\Gamma} = 
\sum\limits_{r=0}^N
\calN_{N,\mu_1\ldots\mu_{r}}
\int\!\rd^D\!\!\momq\; \f{q^{\mu_1}\ldots q^{\mu_{r}}}{ 
D_{0}\!\cdots\! D_{N-1}}
\ee
are either reduced a posteriori, using external 
libraries such as Collier \cite{Denner:2016kdg}, or on the fly, i.e.~during the amplitude construction 
\cite{Buccioni:2017yxi}, with Collier or OneLoop \cite{vanHameren:2010cp} for the final evaluation of scalar integrals. This completely generic algorithm is fully implemented for NLO QCD and NLO EW and available in the public \OpenLoops{} tool \cite{Buccioni:2019sur}.

\section{Structure of a two-loop OpenLoops tool}

A full NNLO calculation consists of a double-virtual, real-virtual and real-real part. The latter two are already provided by the public \OpenLoops{} tool,
as well as the squared one-loop amplitude entering the renormalised double-virtual
contribution,
\be
\calW_{\ssst{NNLO}}^{\ssst{virtual}}\,=\, \f{1}{N_{\rm{hcs}}} \lb
\sum\limits_{\heli,\col} 
2\,\re \Big[\calM_{0}^*(\heli)\,\mathbf{R}\barM_{2}(\heli)\Big] + |\mathbf{R}\barM_{1}(\heli)|^2 \rb,
\label{M2Wtwobar}
\ee
where the bar marks the amplitude in $D$ dimensions and the operator $\mathbf{R}$ the renormalisation procedure. In the following, we focus on the crucial piece for which new efficient methods need to be developed and implemented in the OpenLoops framework, namely the Born two-loop interference.
The numerators of two-loop integrals are again decomposed into a part that can be numerically constructed in four dimensions, and $(D-4)$-dimensional remainders. In
\cite{Pozzorini:2020hkx,Lang:2020nnl,Lang:2021hnw} it was demonstrated that the renormalised $D$-dimensional two-loop amplitude can be split into amplitudes computed with four-dimensional loop numerators,
\be
\mathbf{R}\barM_{2}(\heli) = \calM_{2}(\heli) + \calM_{1,\rm{1l-CT}}(\heli) + \calM_{0,\rm{2l-CT}}(\heli) + \calM_{0,2\times(\rm{1l-CT})}(\heli),
\ee
where the four terms on the rhs are the unrenormalised two-loop amplitude, 
the one-loop amplitude with one-loop rational and UV counterterm insertions,
the tree-level amplitude with two-loop rational and UV counterterm insertions, and the tree-level amplitude with double one-loop rational and UV counterterm insertions.\footnote{The universal two-loop rational terms of UV origin were computed in \cite{Pozzorini:2020hkx,Lang:2020nnl,Lang:2021hnw} for QED, QCD and QCD corrections of the SM for any renormalisation scheme. Potential rational terms originating from the interplay of $(D-4)$-dimensional loop numerator parts and IR divergences are currently under investigation.}
The most challenging part is the Born two-loop interference term constructed with four dimensional numerators
\be
\calW_{\ssst{02}}\,=\, 
\sum\limits_{\heli,\col} 
2\,\re \Big[\calM_{0}^*(\heli)\,\calM_{2}(\heli)\Big] \,=\,
\re\sum\limits_\Gamma
\sum_{\helig}\left(\sum_{\col}2\calM^*_0(\helig)\,\colfac{2}{\Gamma}\right) \amp{2}{\Gamma}(\helig)
,
\label{M2Wtwo}
\ee
where we use \eqref{eq:amp_colfact}, and the sum is taken over the full set of two-loop diagrams $\Gamma$ of the scattering process. 
In the following, we will discuss two-loop diagrams, which become 1PI on amputation of all external subtrees.\footnote{For reducible diagrams we refer to \cite{O2Ldressing}. These diagrams factorise into one-loop contributions, and can be computed with a new algorithm based on the existing one-loop machinery. This is also fully implemented.}

The colour-stripped amplitude $\amp{2}{\Gamma}$ of an irreducible two-loop diagram is constructed from
three chains, $\calCh{1}$, $\calCh{2}$ and $\calCh{3}$, connected by two vertices $\vertex{0}, \vertex{1}$, and has the form
\bea \amp{2}{\Gamma}&=&
\vcenter{\hbox{\scalebox{1.}{\includegraphics[width=0.45\textwidth]{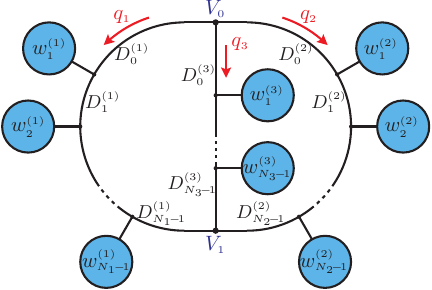}}}} 
=\int\!\rd\momq_1\!\int\!\rd\momq_2
\f{\calN(q_1,q_2)
   }{\prod\limits_{i=1}^3\denc{i}} \Big|_{q_3=-(q_1+q_2)}   \nonumber \\[2mm]
&=& \sum_{r_1=0}^{R_1}\sum_{r_2=0}^{R_2} 
{\calN}_{\mu_1 \cdots \mu_{r_1} \nu_1 \cdots \nu_{r_2}}
\int\!\rd\momq_1\!\int\!\rd\momq_2
\frac{ q_1^{\mu_1}\cdots q_1^{\mu_{r_1}}\,q_2^{\nu_1}\cdots q_2^{\nu_{r_2}}}{\denc{1}\,\denc{2}\,\denc{3}}\Big|_{q_3=-(q_1+q_2)} 
\label{eq:A2d}
\eea
with the three denominator chains $(i=1,2,3)$
\bea
\label{eq:dendef}
\denc{i}&=&
D^{(i)}_0(\bar q_i)\cdots
D^{(i)}_{N_i-1}(\bar q_i)\,,
\qquad\mbox{where}\quad
D^{(i)}_a(\bar q_i) \,=\, \left(\bar q_i + p_{ia}\right)^2-m_{ia}^2\,
\eea
The numerator construction is again performed at the level of tensor coefficients. The tensor integral reduction and evaluation is then the remaining piece to be developed and implemented in our framework.

For the tensor coefficient construction we exploit the factorisation of the numerator into three chains and two connecting vertices,
\be
\calN({ q_1},{ q_2})= \prod\limits_{i=1}^{3}
\numc{i}({ q_i})
\prod\limits_{j=0}^{1}
   \vertex{j}({ q_1},{ q_2}),
\ee
and the factorisation of the chains -- each dependent on a single loop momentum -- into segments of the same structure as the one-loop segments,
\be
\numc{i}(q_i,\{\helisegment{i}{a}\}) = \segment{i}{0}({q_i,\helisegment{i}{0}})\cdots\segment{i}{N_i-1}({q_i},\helisegment{i}{N_i-1}).
\ee
Here we made the dependence of each segment on the helicity d.o.f. $\helisegment{i}{a}$ of the associated subset of external particles explicit.\footnote{The helicity labels are defined in an additive way, such that the global helicity $\helig = \sum\limits_{i=1}^{3} \helic{i}$ is the sum of the chain helicities $\helic{i}=\sum\limits_{a=1}^{N_i-1} \helisegment{i}{a}$, which are constructed from the segment helicities $\helisegment{i}{a}$. For the simplicity of our description in this section, we assume three-point vertices $\vertex{0,1}$. In the case of four-vertices $\vertex{0,1}$ with external subtrees, additional helicity labels are introduced for these two vertices. This is also fully implemented and included in the studies presented in section~\ref{sec:CPUeff}. For details on the helicity definitions and treatment see \cite{O2Ldressing}.}

In order to construct \eqref{M2Wtwo}, the numerator is interfered with the colour factor and full Born amplitude of the process,
\be
\calU({q_1},{q_2}) = \sum\limits_{h} \numpi{-1}({h})\,\calN({q_1},{q_2},{h}), \qquad \numpi{-1}(\helig) = 2\,\sum_{\col}\calM^*_0(\helig)\,\colfac{2}{\Gamma} \label{eq:num_interf}
\ee
The objective is now to construct $\calU({q_1},{q_2})$ in a recursive way, at the level of tensor coefficients, which are then contracted with tensor integrals in the two loop momenta $\momq_1, \momq_2$. In order to find the most efficient recursion with $N_{r}$ steps,
\be
\calV_n = \calV_{n-1} \seg_n \qquad (n=1,\ldots, N_{r})
\ee
with $\calV_{N_{r}}=\calU({q_1},{q_2})$ and the building blocks $\seg_n \in \{\seg^{(i)}_a,\vertex{0,1},\numc{i},\numpi{-1}\}$,
a CPU cost analysis for the possible algorithms of this form was performed, each for several QED and QCD Feynman diagrams. Here we estimated the CPU cost of each step by the number of multiplications, the most expensive numerical operation. The most efficient recursion was then fully implemented and validated for QED and QCD corrections to SM processes. It consists of the following steps. 
\begin{itemize}[leftmargin=6mm]
 \item[0.] The three chains are sorted by their number of segments, such that $N_1 \geq N_2 \geq N_3$. The order of $\vertex{0}$ and $\vertex{1}$ is determined by vertex type,
 such that the number of multiplications in the following steps is minimal (for details see \cite{O2Ldressing}).
 \item[1.] The shortest chain $\calCh{3}$ is constructed through the recursion
 \be
 \numpc{3}{n}({ q_3},{ \helipc{3}{n}}) = \numpc{3}{n-1}({ q_3},{ \helipc{3}{n-1}})\cdot\segment{3}{n}({ q_3},{ \helisegment{3}{n}}) 
 \quad\text{with} \quad
 \helipc{3}{n} = \sum\limits_{a=1}^{n} \helisegment{3}{a}
 \ee
 and $n=0,\ldots,N_3-1$. This is usually the chain with the least helicity d.o.f. and intermediate results can be recycled in multiple Feynman diagrams, such that this step is negligible in the overall CPU cost for a full process.
 \item[2.] 
 The full diagram is then constructed through a sequence of sub-recursions:
 \begin{itemize}[leftmargin=3mm]
    \item[2.1] The longest chain $\calCh{1}$ is constructed through steps
\be    
\numpi{n}({ q_1},{ \helipcc{1}{n}})
=
{\sum\limits_{\helisegment{1}{n}}}
\numpi{n-1}({ q_1},{ \helipcc{1}{n-1}})\cdot\segment{1}{n}({ q_1},{ \helisegment{1}{n}})
\quad 
\text{with}\quad
{ \helipcc{1}{n}= \helig -\sum\limits_{a=1}^n\helisegment{1}{a}}
\ee
and $n=0,\ldots,N_1-1$. The initial condition defined in \eqref{eq:num_interf} contains the interference with the full Born, which allows for the on-the-fly summation of the helicities of each chain segment during the recursion step, in which it is attached.\footnote{The on-the-fly summation of helicities was already introduced in \cite{Buccioni:2017yxi} for the one-loop algorithm.} Since $\calCh{1}$ is the longest chain, a large portion of helicity d.o.f is already summed over at a stage at which the partially constructed diagram depends only on a single loop momentum.
    \item[2.2] The two-loop vertex $\vertex{1}$ is connected to the previously constructed chains,
\be
\numinter({ q_1},{ q_3},{ \helic{2}}) 
=
{ \sum\limits_{\helic{3}}}\;
\numpi{N_1-1}({ q_1},{ \helig-\helic{1}})\;
\numpc{3}{N_3-1}({ q_3},{ \helic{3}})\;
\vertex{1}({ q_1},{ q_3})
\ee
summing over the helicities of $\calCh{3}$, and introducing the dependence on a second loop momentum, and hence a much higher complexity.
    \item[2.3] The two-loop vertex $\vertex{0}$ is connected,
\be
\numpd{-1}({ q_1},{ q_2},{ \helic{2}})
= 
\numinter({ q_1},{ q_3},{ \helic{2}})\;
\vertex{0}({ q_1},{ q_2})\Big|_{ q_3= -(q_1+q_2)}
\ee
which reduces the number of open Lorentz/spinor indices from three to two.
    \item[2.4] The remaining chain $\calCh{2}$ is constructed through steps
\be
\numpd{n}({ q_1},{ q_2},{ \helipccloc{2}{n}}) 
=
{ \sum\limits_{\helisegment{2}{n}}}
  \numpd{n-1}({ q_1},{ q_2},{ \helipccloc{2}{n-1}}) \; \segment{2}{n}({ q_2},{ \helisegment{2}{n}})
  \quad 
\text{with}\quad
  \helipccloc{2}{n}=\sum\limits_{a=n+1}^{N_2-1} \helisegment{2}{a}
\ee
and $n=0,\ldots,N_2-1$.
Here the complexity stemming from the high tensor ranks in the loop momenta is counterbalanced by the dependence on only a few remaining helicities. 
By construction, in the final result \be \calU({q_1},{q_2}) = \numpd{N_2-1}({ q_1},{ q_2},0) \ee all helicities are summed.
 \end{itemize}
\end{itemize}
This algorithm is completely generic. For QED and QCD corrections to the SM, 
it has been fully implemented and validated at the level of tensor coefficients in the OpenLoops framework.

\section{CPU efficiency and numerical stability} \label{sec:CPUeff}

\begin{figure}[t]
\centering
\includegraphics[trim = 5mm 1mm 5mm 14mm, clip, height=78.5mm]{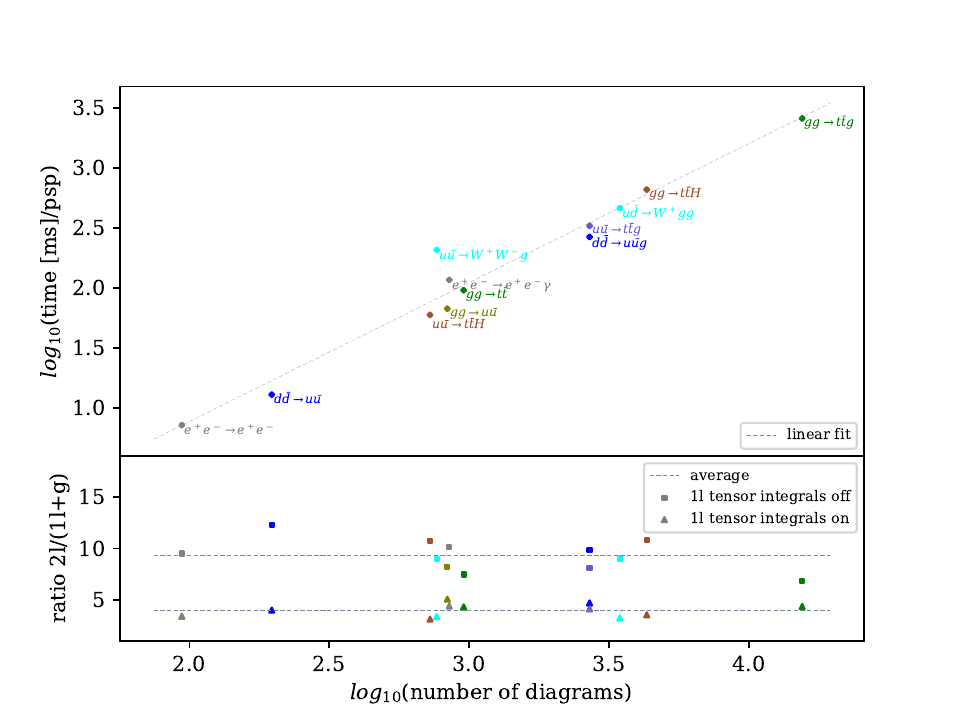}
\caption{Computation time for QCD and QED corrections to selected SM processes. For processes with external $e^{\pm}$, QED corrections were considered, for all other processes QCD corrections. Amplitudes with external tops were computed with a massive top, all others with purely massless internal fermions.}
\label{fig:speed}
\end{figure}

In order to test the CPU efficiency of this new algorithm, we computed the tensor coefficients for a wide range of QED and QCD 
processes,
each for $1000$ uniform random phase space points (psp) on a computer with a
single Intel i7-6600U @ 2.6 GHz processor and 16GB RAM. The average time per psp is shown in the upper plot of Fig.~\ref{fig:speed} against the number of Feynman diagrams. The runtimes for the complete two-loop recursion, including full colour and helicity sums, range from a few ms for simple QED and QCD processes to $\mathcal{O}(1s)$ for more complex $2 \to 3$ processes. The computation time scales linearly with the number of diagrams.
It is interesting to compare these two-loop (2l) time measurements to the ones for the corresponding real-virtual corrections, \ie the same process with one additional photon or gluon at one-loop level (1l+g), which constitutes another, already fully available component of a full NNLO calculation. The ratio of 2l and 1l+g timings is depicted in the lower plot of Fig.~\ref{fig:speed}, once including
only the tensor coefficient constructions, and once including the tensor integrals in the one-loop calculation. These ratios are fairly constant over all considered processes with
\beas 
\frac{\text{2l (tensor coefficients)}}{\text{ 1l+g (tensor coefficients)}} &=& 9 \pm 3, \qquad
\frac{\text{2l (tensor coefficients)}}{\text{ 1l+g (full calculation)}} = 4 \pm 1{}.
\eeas
Compared to the corresponding one-loop tensor coefficient construction with two extra gluons/photons, the  
two-loop tensor integral construction is even a factor $3-8$ faster (see \cite{O2Ldressing}).
Considering the much higher complexity of two-loop diagrams as compared to one-loop diagrams, these are very promising values,
and we expect that
the efficiency of the future tool for full two-loop calculations will largely depend on the efficiency of the tensor reduction.

Our implementation of this new algorithm also shows high numerical stability at the level of the tensor integral coefficients, as demonstrated by relative uncertainty measurements for $2\to 2$ and $2\to 3$ QCD amplitudes computed in double precision for $10^5$ uniform random psp. These relative uncertainties are in the range of $10^{-16}$ to $10^{-14}$ for the bulk of the psp, and never below order $10^{-12}$ and $10^{-11}$. For details we refer to \cite{O2Ldressing}.
\section{Conclusion}
We presented a completely new algorithm for the CPU efficient and numerically stable construction of the loop-momentum tensor coefficients of two-loop amplitudes. This is an important building block in the development of a fully automated two-loop tool in the \OpenLoops{} framework.
\section*{Acknowledgements}
This research was supported by the Swiss National Science Foundation (SNSF) 
under the SNSF Ambizione grant PZ00P2-179877. The work of S.P. 
was supported through contract BSCGI0-157722.


\nolinenumbers

\end{document}

%% file: macros_OL_2l.tex
\newcommand{\OpenLoops}{\text{OpenLoops}}

\newcommand{\ie}{i.e.\ }

\newcommand{\f}[2]{\frac{#1}{#2}}

\newcommand{\ssst}[1]{\scriptscriptstyle{\text{#1}}}
\newcommand{\nosss}[1]{#1}

\newcommand{\bea}{\begin{eqnarray}}
\newcommand{\eea}{\end{eqnarray}}
\newcommand{\be}{\begin{equation}}
\newcommand{\ee}{\end{equation}}
\newcommand{\ba}{\begin{align}}
\newcommand{\ea}{\end{align}}
\newcommand{\beas}{\begin{eqnarray*}}
\newcommand{\eeas}{\end{eqnarray*}}
\newcommand{\bes}{\begin{equation*}}
\newcommand{\ees}{\end{equation*}}
\newcommand{\bas}{\begin{align*}}
\newcommand{\eas}{\end{align*}}

\newcommand{\eps}{{\varepsilon}}

\newcommand{\lb}{\left(}
\newcommand{\rb}{\right)}

\newcommand{\momk}[1]{k_{\nosss{#1}}}

\newcommand{\mass}[1]{m_{\nosss{#1}}}
\newcommand{\heli}{h}

\newcommand{\helicheck}{\check h}

\newcommand{\helihat}{\hat h}

\newcommand{\momq}{\bar{q}}

\newcommand{\calA}{\mathcal{A}}

\newcommand{\calM}{\mathcal{M}}
\newcommand{\calN}{\mathcal{N}}

\newcommand{\barM}{\bar{\mathcal{M}}}

\newcommand{\calV}{\mathcal{V}}
\newcommand{\calW}{\mathcal{W}}
\newcommand{\seg}{S}

\newcommand{\calU}{\mathcal{U}}
\newcommand{\helicheckloc}{\tilde h}

\newcommand{\segment}[2]{\seg_{#2}^{(#1)}}

\newcommand{\helisegment}[2]{\heli_{#2}^{(#1)}}
\newcommand{\helic}[1]{\heli^{(#1)}}                   
\newcommand{\helig}{\heli}                             
\newcommand{\helipc}[2]{\helihat_{#2}^{(#1)}}          
\newcommand{\helipcc}[2]{\helicheck_{#2}^{(#1)}}       
\newcommand{\helipccloc}[2]{\helicheckloc_{#2}^{(#1)}} 

\newcommand{\numc}[1]{\calN^{(#1)}}                 
\newcommand{\numpc}[2]{\calN^{(#1)}_{#2}}           
\newcommand{\numpi}[1]{\calU^{(1)}_{#1}}           
\newcommand{\numpd}[1]{\calU^{(123)}_{#1}}           
\newcommand{\numinter}{{\calU^{(13)}}}                

\newcommand{\amp}[2]{{\calA}_{{#1},{#2}}}

\newcommand{\fullamp}[2]{{\calM}_{{#1},{#2}}}

\newcommand{\colfac}[2]{{C}_{{#1},{#2}}}
\newcommand{\vertex}[1]{\calV_{#1}}
\newcommand{\calCh}[1]{\mathcal{C}_{#1}(\bar q_{#1})}  
\newcommand{\denc}[1]{\mathcal{D}^{(#1)}(\bar q_{#1})} 

\newcommand{\col}{\mathrm{col}}

\newcommand{\re}{\mathrm{Re}}

\newcommand{\rd}{\mathrm d}

\definecolor{bluemar}{rgb}{0,0,.5}
\definecolor{redmar}{rgb}{.8,0,0}
\definecolor{greenmar}{rgb}{0,.5,0}